\documentclass{jpp}

\usepackage{natbib}
\usepackage{graphicx}
\usepackage{upmath}
\usepackage{amssymb}

\usepackage{xcolor}
\usepackage{epstopdf}

\def\beq{\begin{equation}}
\def\eeq{\end{equation}}
\def\een#1{\label{#1} \end{equation}}
\def\beqa{\begin{eqnarray}}
\def\eeqa{\end{eqnarray}}
\def\ean#1{\label{#1} \end{eqnarray}}
\def\pd#1#2{\frac{\partial{#1}}{\partial{#2}}}

\def\eqref#1{(\ref{#1})}
\def\nnn{\nonumber \\}
\def\eps{\varepsilon}
\def\etal{\textit{et al.\/}}
\def\sech{{\rm sech}}

\begin{document}

\title[Soliton and electric field collisions in critical plasmas]%
    {Collisions of acoustic solitons and their electric fields in plasmas at
        critical compositions}
\author[F. Verheest and W. A. Hereman]
{F\ls R\ls A\ls N\ls K\ns V\ls E\ls R\ls H\ls E\ls E\ls S\ls T $^{1,2}$
    \footnote{Email address for correspondence: frank.verheest@ugent.be} and
W\ls I\ls L\ls L\ls Y\ns A.\ns H\ls E\ls R\ls E\ls M\ls A\ls N$^3$
    \footnote{Email address: whereman@mines.edu} }
\pubyear{2018}
\affiliation{$^1$Sterrenkundig Observatorium, Universiteit Gent,
    Krijgslaan 281, B--9000 Gent, Belgium \\[\affilskip]
$^2$School of Chemistry and Physics, University of KwaZulu-Natal,
        Durban 4000, South Africa \\[\affilskip]
$^3$Department of Applied Mathematics and Statistics, Colorado School of Mines,
        Golden, Colorado 80401-1887, USA }



\maketitle

\begin{abstract}
Acoustic solitons obtained through a reductive perturbation scheme are normally
governed by a Korteweg-de Vries (KdV) equation.
In multispecies plasmas at critical compositions the coefficient of the
quadratic nonlinearity vanishes.
Extending the analytic treatment then leads to a modified KdV (mKdV) equation,
which is characterized by a cubic nonlinearity and is even in the electrostatic
potential.
The mKdV equation admits solitons having opposite electrostatic polarities, in
contrast to KdV solitons which can only be of one polarity at a time.
A Hirota formalism has been used to derive the two-soliton solution.
That solution covers not only the interaction of same-polarity solitons but
also the collision of compressive and rarefactive solitons.
For the visualisation of the solutions, the focus is on the details of the
interaction region.
A novel and detailed discussion is included of typical electric field
signatures that are often observed in ionospheric and magnetospheric plasmas.
It is argued that these signatures can be attributed to solitons and their
interactions.
As such, they have received little attention.
\end{abstract}

\section{Introduction}

Acoustic solitons in plasmas, obtained through a reductive perturbation theory
(RPT), are normally governed by a Korteweg-de Vries equation.
The equation was originally derived to model solitary waves observed on the
surface of shallow water [\cite{KdV,AblowitzClarkson1991}], but was seventy
years later found to have applications in various other fields of physics,
notably in plasmas [\cite{Gardner1967,Gardner1974}].
Because of the algorithmic nature of RPT it became a popular tool in
theoretical studies of nonlinear acoustic waves in plasmas, for a variety of
different model compositions, resulting in a large number of scholarly
publications.

It was also established that equations such as the KdV equation are completely
integrable, in the sense that besides the single soliton solution there are
also $N$-soliton solutions (for any positive integer $N$) which collide
elastically.
Even the one-soliton solution shows that there is an intricate relation between
the amplitude, width and velocity of the wave.
In particular, larger solitons are narrower but travel faster.
For a typical two-soliton interaction this means that a later-launched larger
soliton will overtake a slower one.
After an intricate collision, both emerge unscathed except for small phase
shifts.
Hence the name ``soliton" was coined \cite{Zabusky} for the resemblance with
interacting particles.

The colloquial term ``later-launched" has to be understood as follows.
A two-soliton solution is not merely a superposition of two single solitons, as
this would be meaningless for a \textit{nonlinear} equation.
Rather, a two-soliton solution has a complicated nonlinear structure which
asymptotically reduces to two separated single solitons.
During the collision the nonlinearity affects their individual profiles in
quite unexpected ways.
However, after the collision the two solitons again separate, and resume their
initial shapes as if nothing has happened, except for small phase shifts.
Indeed, the faster soliton is shifted forward while the slower soliton is
shifted backward, relative to the positions where individual solitons would
have been had they not collided.

When the plasma model allows for changes in polarities for critical values of
the compositional parameters [\cite{DasGC,DasTagare,Tagare,Lee,Saini}], the
analytic treatment then leads to a modified Korteweg-de Vries (mKdV) equation
with cubic nonlinearity.
The equation is even in the electrostatic potential $\varphi$, thus admitting
solitons of opposite electrostatic polarities.
In contrast, KdV solitons can only be of one polarity at a time.
Recall that the historic application of KdV theory to surface waves on shallow
water yielded only compressive solitons, in the shape of humps.
Water surfaces do not sustain rarefactive solitons, also called holes or dips.

To study the overtaking interactions between two solitons, Hirota's bilinear
method has been applied, which can deal with any number of interacting waves
because the KdV and mKdV equations are completely integrable.
Our visualization of two soliton interactions will be focussed on what happens
at the center of the interaction region, not only for the electrostatic
solitons but also for their derivatives, which yield the electric field
profiles.
To the best of our knowledge, electric field profiles have not been studied
before.

As an aside, there is no Hirota or equivalent formalism that leads to solutions
describing head-on collisions, neither for KdV nor mKdV equations.
What is available in the literature relies on an extension of the
Poincar\'e-Lighthill-Kuo (PLK) formalism of strained coordinates, which leads
to approximate results of limited use [\cite{HeadOnCairns,HeadOnIons}].
The shortcoming of the PLK method is that it uses an addition of the amplitudes
in an essentially nonlinear problem.
Furthermore, its results contradict recent laboratory experiments
[\cite{Harvey}] and numerical simulations [\cite{Kakad2017,Kumar}].

In Section 2 we recall the essentials of RPT leading to mKdV equations,
together with their one-soliton solutions, well studied in the plasma and
mathematical physics literature.
We summarize the steps of the Hirota bilinear method for the mKdV equation in
Section 3.
The visualization of these interactions is covered in Section 4, first for a
collision of two humps, next for a hump and a hole.
Novel features of these interactions, in particular for electric fields, are
discussed in Section 5.
Electric fields are often seen in ionospheric and magnetospheric satellite
observations, but their overtaking interactions have not been studied before.
The paper concludes with a brief summary of the results in Section 6.

\section{Reductive perturbation formalism and evolution equations}

There is a plethora of plasma compositions treated in the literature, allowing
critical densities leading to the mKdV equation.
These plasmas are usually described by a number of cold and/or warm fluid
species, of quite different characteristics, in the presence of some
inertialess Boltzmann or nonthermal distributions for the hot species.

RPT rests on two pillars: a suitable stretching of space and time and the
expansion of the dependent variables.
For the family of KdV-type equations the stretching can be chosen as
\begin{equation}
\xi = \eps (X - V T), \qquad\qquad \tau = \eps^3 T,
\label{stretch}
\end{equation}
or equivalent choices, with $V$ the linear acoustic phase velocity, $\eps$ a
small parameter and $X$ and $T$ the physical space and time coordinates,
respectively.
This is inspired by the dispersion law for linear waves with frequency $\omega$
and wave number $k$, by taking the limit $k\to 0$ but keeping the phase
velocity $\omega/k$ finite, based on the idea that acoustic modes have $\omega$
proportional $k$ to lowest order.
For the expansion of the electrostatic potential, $\varphi,$ we take
\begin{equation}
\varphi = \eps \varphi_1 + \eps^2 \varphi_2  + \eps^3 \varphi_3 + \ldots,
\label{expan}
\end{equation}
with similar expansions for the plasma variables like densities, pressures and
fluid velocities of the different species.

Applying \eqref{stretch} and \eqref{expan} to the basic fluid equations and
Poisson's equation produces order by order in $\eps$ sets of equations for the
successive $\varphi_i$.
In the generic case, $\varphi_1 \equiv 0$ leads to the KdV equation,
\begin{equation}
A \pd{\varphi_2}{\tau} + B \varphi_2\, \pd{\varphi_2}{\xi}
    + \frac{1}{2}\, \pd{^3 \varphi_2}{\xi^3} = 0.
\label{KdV}
\end{equation}
The coefficients $A$ and $B$ include the compositional plasma parameters and
$V$; the latter as a solution of the linear dispersion law is a function of
those same parameters.
No matter how complicated the plasma model is, one gets the KdV equation
\eqref{KdV} which has soliton solutions and other special properties, provided
the plasma species obey barotropic pressure-density relations
[\cite{VerheestASSL}].
Given its ubiquity in physics, this is the best known and most studied
nonlinear soliton.

We assume that the plasma composition is critical ($B=0$).
Continuing with $\varphi_1\not = 0$ then leads to the mKdV equation,
\begin{equation}
A \pd{\varphi_1}{\tau} + C \varphi_1^2\, \pd{\varphi_1}{\xi}
    + \frac{1}{2}\, \pd{^3 \varphi_1}{\xi^3} = 0,
\label{mKdV}
\end{equation}
where $A$ and $C$ again depend on the plasma parameters and $V$.
Perusing the literature, it becomes clear that almost all plasmas need to have
at least three constituents before $B=0$ can be imposed under the condition
that $V$ annuls the linear dispersion relation.

The earliest examples are plasmas with two cold ion species (one positive, one
negative) in the presence of Boltzmann electrons
[\cite{DasGC,DasTagare,Watanabe,Tagare}].
Later, this was extended to include more cold or warm ion species
[\cite{TagareCrit}], and the picture remains unchanged when various nonthermal
distributions are introduced for the hot electrons [\cite{TagareKdV}], such as
the kappa distribution [\cite{Vasyliunas,SumThor}] and Tsallis nonextensive
distribution [\cite{Tsallis,Lima2000}].
The only exception to this three or more species rule is the Cairns
distribution [\cite{Cairns}] because it admits a critical density in the
presence of only one cold ion species.
Regardless of the finetuning of the parameters needed for criticality, mKdV
solitons were even observed experimentally in multi-ion plasmas and their
collision properties were investigated by numerically solving the mKdV equation
[\cite{Nakamura}].

Note that the mKdV equation \eqref{mKdV} is invariant under the change
$\varphi$ to $-\varphi$.
As a consequence of this uncommon property, every positive polarity soliton has
an equivalent negative one.
The one-soliton solution of \eqref{mKdV} is well known,
\begin{equation}
\varphi_1 = \pm \sqrt{\frac{6 U\! A}{C}}\, \sech
    \left[ \sqrt{2 U\! A} (\xi - U \tau) \right],
\label{mKdVSoliton}
\end{equation}
where $U$ is an arbitrary velocity, in the frame moving with velocity $V$ with
respect to an inertial observer.
On the other hand, the two-soliton solution is much less known in the context
of plasma physics, and will therefore be the main focus of our efforts in the
present paper.

In what follows, we will start from a generic form with coefficients chosen
such that the application of Hirota's formalism
[\cite{HirotaKdV,HirotamKdV,Hirota}] is as simple as possible [\cite{Drazin}].
A change of variables,
\begin{equation}
\xi = 2 x\,\sqrt{\frac{3}{C}}, \qquad
\tau = 48 t\,\frac{A}{C}\,\sqrt{\frac{3}{C}}, \qquad \varphi_1 = u,
\end{equation}
transforms the mKdV equation \eqref{mKdV} into
\begin{equation}
u_t + 24 u^2 u_x + u_{xxx} = 0. \label{mKdVh}
\end{equation}
This transformation requires that $C>0$ (and $B=0$).
The coefficient 24 in \eqref{mKdVh} is chosen to minimize the numerical factors
when applying Hirota's method, and derivatives are denoted by lower-case
subscripts $x$ and $t$, where, for example, $u_{xxx}$ denotes the third
derivative of $u$ with respect to $x$.
Note that we have transformed the space and time variables, but $u$ remains a
normalized electrostatic potential at the lowest nonzero order.

\section{Hirota's bilinear method and soliton solutions}

\cite{HirotaKdV} developed an ingenious method to find exact $N$-soliton
solutions for the KdV equation.
The method uses bilinear operators, hence the name \textit{Hirota's bilinear
method}.
It was later shown that the method can be applied to large classes of nonlinear
evolution equations, including the mKdV equation.

The key idea is to change the dependent variable so that the given nonlinear
equation becomes bilinear in one or more new dependent variables.
Once the appropriate bilinear forms have been found, a formal series expansion
is used to generate its solutions in an iterative way.
If pure solitons exist the iterative process terminates at a certain level and
the finite series leads to an exact solution.

Hirota's method proceeds in three steps:
(i) a judicious guess for the transformation of the dependent variable,
(ii) writing the transformed equation as a single bilinear equation or a
coupled system of bilinear equations, and
(iii) using a formal expansion scheme to solve these bilinear equation(s).

Inspiration for the initial step has sometimes come from performing the
Painlev\'e integrability test [\cite{AblowitzClarkson1991,Drazin}] of an
evolution equation or knowing its $N$-soliton solution from application of the
inverse scattering method.

Let us now recall some of the steps for the mKdV equation.
\cite{HirotaKdV} introduced bilinear differential operators, $D_x$ and $D_t$,
defined for ordered pairs of arbitrary functions $f(x,t)$ and $g(x,t)$, as
follows,
\begin{equation}
D_x^m D_t^n  \{f \mathbf{\cdot } g \}
= \left( \pd{}{x} - \pd{}{x'} \right)^m
  \left( \pd{}{t} - \pd{}{t'} \right)^n
    f(x,t) g(x',t') \bigg|_{x'=x, t'=t}\, ,
\end{equation}
where $m$ and $n$ are nonnegative integers.

Operators like $D_x$ and $D_t$ are bilinear because of their evident linearity
in both arguments $f$ and $g$.

Whereas the KdV equation can be replaced by one bilinear equation
[\cite{HirotaKdV,Drazin}], the mKdV equation \eqref{mKdVh} requires a coupled
system of bilinear equations [\cite{HirotamKdV}] because the change of the
dependent variable
\begin{equation}
u = \frac{\partial}{\partial x} \Big( \arctan \left( \frac{f}{g} \right) \Big)
= \frac{f_x g - f g_x}{f^2 + g^2}
\label{mKdV-Hirota}
\end{equation}
involves two functions $f(x,t)$ and $g(x,t)$.
Substitution of \eqref{mKdV-Hirota} into \eqref{mKdVh} yields, after one
integration with respect to $x$,
\begin{eqnarray}
&& (f^2 + g^2)(f_t g - f g_t + f_{xxx} g - 3 f_{xx} g_x + 3 f_x g_{xx}
    - f g_{xxx}) \nnn
&& - 6 (f g_x - f_x g) (f\!\,f_{xx} - f_x^2 + g\!\,g_{xx} - g_x^2) = 0.
\label{mKdV-infg}
\end{eqnarray}
Setting each term equal to zero results in the following pair of bilinear
equations,
\begin{eqnarray}
&& (D_t + D_x^3) \{f \mathbf{\cdot} g\} = 0, \label{B1} \\
&& D_x^2 \{f \mathbf{\cdot} f + g \mathbf{\cdot} g\} = 0. \label{B2}
\end{eqnarray}
The expansions are
\begin{eqnarray}
f &=& f_0 + \eps f_1 + \eps^2 f_2 + \ldots, \nnn
g &=& g_0 + \eps g_1 + \eps^2 g_2 + \ldots,
\label{fgexpansion}
\end{eqnarray}
where $f_0$ and $g_0$ are constants (not both zero to avoid a trivial solution)
and $\eps$ is a bookkeeping parameter to disentangle the different orders.
Once the successive $f_i$ and $g_i$ have been computed, one sets $\eps = 1$.
Doing the computations, one finds from the lowest nonzero order of \eqref{B2}
that either $f_0 = g_1 = 0$ or $g_0 = f_1 = 0$.
Both choices are equivalent for they amount to a sign change in the polarity of
the nonlinear wave.

To derive the one and two-soliton solutions, we continue with $f_0 = g_1 = 0$.
Then, without loss of generality, we normalize $g_0=1$ and recover from
$f_1 = {\rm e}^{\theta_1}$ (in the one-soliton case) the well-known sech
solution \eqref{mKdVSoliton}.
Here and below one uses the notation $\theta_i = k_i x - k_i^3 t + \delta_i$,
which incorporates the linear dispersion $\omega_i = k_i^3$ characterizing the
KdV-family.
In the two-soliton case, $f_1 = {\rm e}^{\theta_1} + {\rm e}^{\theta_2}$ and
one can set $f_2 = 0$.
For one and two-soliton interactions one can shift the origins of $x$ and $t$
so as to absorb $\delta_i$, which will henceforth be omitted.
The same argument does no longer work for three-soliton interactions (and
higher).
Suppressing the $\delta_i$ then requires non-unit amplitudes
$a_i = {\rm e}^{\delta_i}.$

After computing $g_2$ and verifying that $g_i = 0$ for $i \geqslant 3$ and
$f_i = 0$ for $i \geqslant 2$, one obtains
\begin{eqnarray}
 \nnn
f &=& {\rm e}^{\theta_1} + {\rm e}^{\theta_2}, \nnn
g &=& 1 - \frac{(k_1-k_2)^2}{(k_1+k_2)^2}\, {\rm e}^{\theta_1 + \theta_2},
\end{eqnarray}
and, from \eqref{mKdV-Hirota},
\begin{equation}
u = \frac{k_1 {\rm e}^{\theta_1} + k_2 {\rm e}^{\theta_2}
  + \displaystyle{\frac{(k_1-k_2)^2}{(k_1+k_2)^2}}
    (k_1 {\rm e}^{\theta_2} + k_2 {\rm e}^{\theta_1})
        {\rm e}^{\theta_1 + \theta_2}}{
    1 + {\rm e}^{2\theta_1} + {\rm e}^{2\theta_2}
  + \displaystyle{\frac{8 k_1 k_2}{(k_1 + k_2)^2}}\, {\rm e}^{\theta_1+\theta_2}
  + \displaystyle{\frac{(k_1-k_2)^4}{(k_1+k_2)^4}}\,
    {\rm e}^{2\theta_1+2\theta_2}}.
\label{umKdVHir}
\end{equation}
This is a family of two-soliton solutions of \eqref{mKdV-Hirota} with arbitrary
constants $k_1$ and $k_2$.
Written as such, it is not readily imagined what the solution looks like.
To get an idea of the shape one has to resort to visualizations which will be
the focus of the next section.
We have also checked that \eqref{umKdVHir} solves the mKdV equation
\eqref{mKdVh}, as other (equivalent) expressions of the solution are possible
[\cite{Anco}].

\section{Interaction dynamics of two solitons for the mKdV equation}

In this section we will visualize the dynamics of the two-soliton solution
\eqref{umKdVHir}.
The wave numbers $k_1$ and $k_2$ determine the amplitude and polarity of the
two solitons comprised inside \eqref{umKdVHir}, and through the linear
dispersion, also the phase speed $k_i^2$.
Far away from the collision region, the two solitons each assume the shape of a
single soliton \eqref{mKdVSoliton}.
The larger the wave number $k_i$, the taller the soliton will be and the faster
it will travel since the phase velocity equals $k_i^2$.
A positive wave number produces a hump (or bump); a negative wave number
produces a hole (sometimes referred to as dip or depression), but both travel
to the right.

To visualize the asymptotic behaviour (in particular, separation into
individual solitons) we use long time intervals in Figures \ref{CC} and
\ref{CR}, typically for $t$ varying between $-40$ and $40$, with large (equal)
time steps, e.g., $\Delta t = 40$.
To zoom into the collision regions, shown in Figures \ref{CCcent} and
\ref{CRcent}, we let $t$ vary between, say, $-5$ and $5$, and use small (equal)
time steps of length 5.
All pictures show snapshots of both solitons at various moments in time to
locate them on the $x$-axis before, during, and after their collision, going
from left to right.
The large soliton peaks are easily identified, far away from the interaction
region.

\subsection{Overtaking of same polarity solitons}

\begin{figure}
\centerline{
\includegraphics[width=100mm]{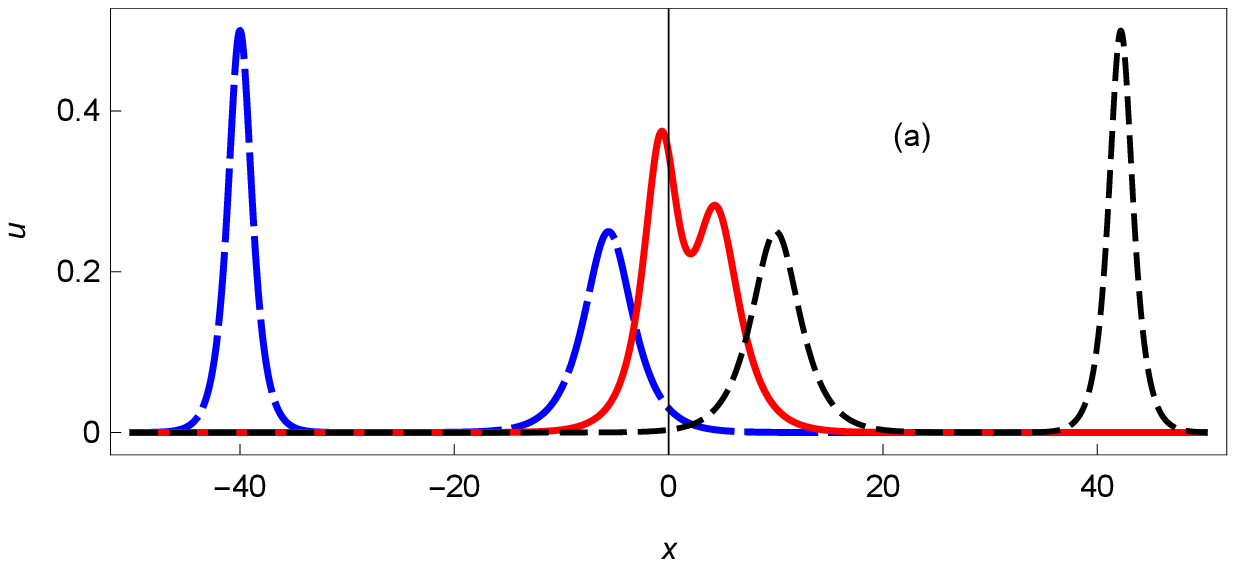}}
\centerline{
\includegraphics[width=100mm]{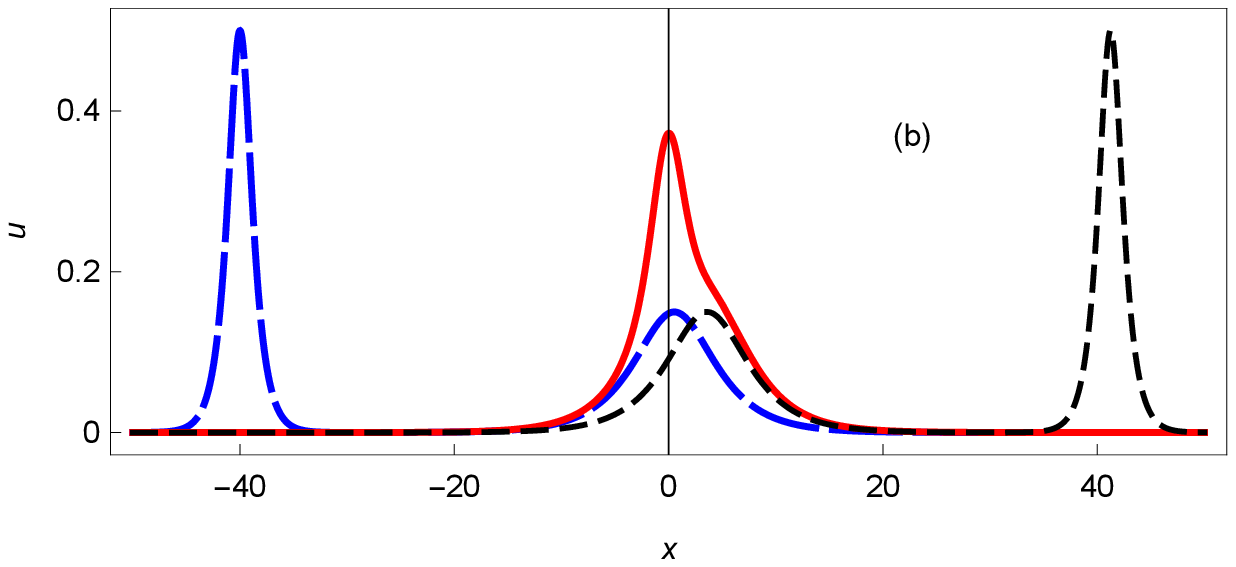}}
\caption{
Typical overtaking of a smaller by a larger hump, for (a) $k_1=0.5$ and $k_2=1$
and (b) $k_1=0.3$ and $k_2=1$.
The curves correspond to $t=-40$ (blue dashed), 0 (red) and 40 (black dotted),
where the peaks move to the right as time increases.}
\label{CC}
\end{figure}

\begin{figure}
\centerline{
\includegraphics[width=60mm]{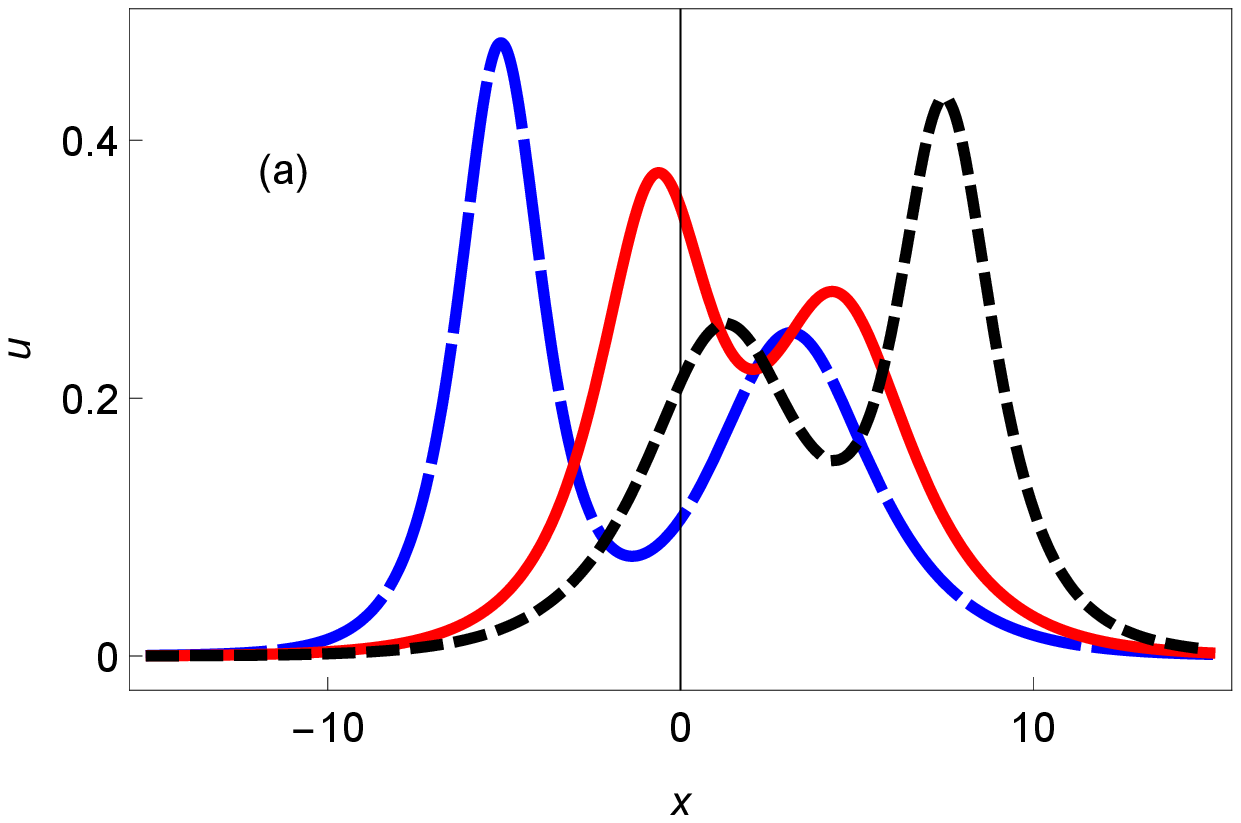} \qquad
\includegraphics[width=60mm]{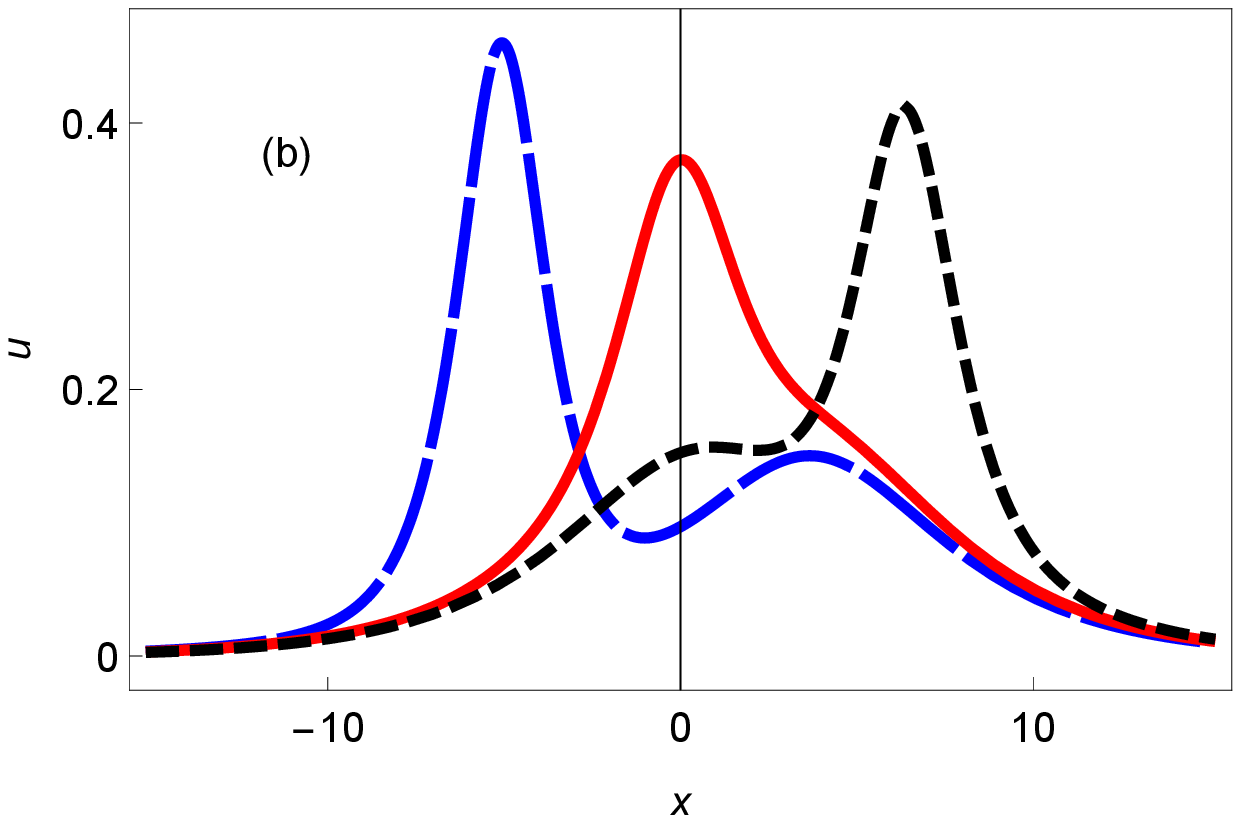}}
\caption{
Details of the overtaking interaction of a smaller by a larger hump, for the
parameters used in Figure \ref{CC}.
The curves correspond to $t=-5$ (blue dashed), 0 (red) and 5 (black dotted).}
\label{CCcent}
\end{figure}

Figure \ref{CC} illustrate the dynamics of solitons with the same polarity (two
humps).
The parameters determining the respective shape of the solitons are (a)
$k_1=0.5$ or (b) $k_1=0.3$, and $k_2=1$.
The curves are shown for $t=-40$ (blue dashed), 0 (red) and 40 (black dotted),
where the peaks move to the right as time increases.
The pictures for two holes would be the same apart from a vertical flip over
the $x$-axis.
The distinguishing feature between parts (a) and (b) of Figure \ref{CC} is what
happens in the interaction zone, details of which are shown in the
corresponding graphs in Figure \ref{CCcent}.

Although we have graphed the solitons at equal time intervals, it can be seen
that when the larger hump has overtaken the smaller one, the net effect is a
slight forward shift of the larger soliton.
Such shifts do not persist in time, and are only to be compared to what the
trajectory of the larger hump would have been, had it been a true one-soliton
solution, alone in the physical system.
After a while the solitons separate and resume their original shapes.
Asymptotically, e.g., for times $t<-40$ or $t>40$, the profile of each soliton
becomes undistinguishable from the one-soliton shape.

To start, Figure \ref{CC} illustrates how a faster, larger but narrower hump
with positive polarity overtakes a slower, smaller but wider one, also with
positive polarity, but for different ratios $k_1/k_2$.
By reducing the time step near the overtaking region we show in Figure
\ref{CCcent} how the interaction evolves between the larger and the smaller
soliton, for the cases shown in Figure \ref{CC}.
The parameters match those in Figure \ref{CC}, but with smaller time steps,
$t=-5$ (blue dashed), 0 (red) and 5 (black dotted).

There are two distinct possibilities.
For the first possibility the parameters are chosen such that during the
interaction there are at all times two distinct peaks, as shown in Figure
\ref{CCcent}(a).
Thus, as the first peak increases in amplitude and the second peak decreases,
it looks as if the larger and the smaller peaks have swapped places.
In other words, after the collision the larger soliton is ahead, having
undergone a forward phase shift, and \textit{mutatis mutandis} for the smaller
soliton.

The other possibility is that, as we decrease $k_1$, keeping $k_2$ fixed, the
figure changes qualitatively and the two solitons temporarily merge into a
single but distorted peak, illustrated in Figure \ref{CCcent}(b).
This occurs for $k_1/k_2<0.39$.
Indeed, only the ratio is important, as we can, without loss of generality,
take $k_2=1$ for the largest of the peaks, and $k_1<1$ for the smaller one.
Analogous results have been found in the mathematics literature [\cite{Anco}].

Switching the polarities from positive to negative produces the same type of
figures, except that the Figures \ref{CC} and \ref{CCcent} are vertically
flipped (not shown).
All this is reminiscent of what happens for the interaction of KdV solitons,
which are always of the same polarity, depending on the sign of $B$ (not shown
here, but presented in many papers in the literature).

\subsection{Overtaking of opposite polarity solitons}

\begin{figure}
\centerline{
\includegraphics[width=100mm]{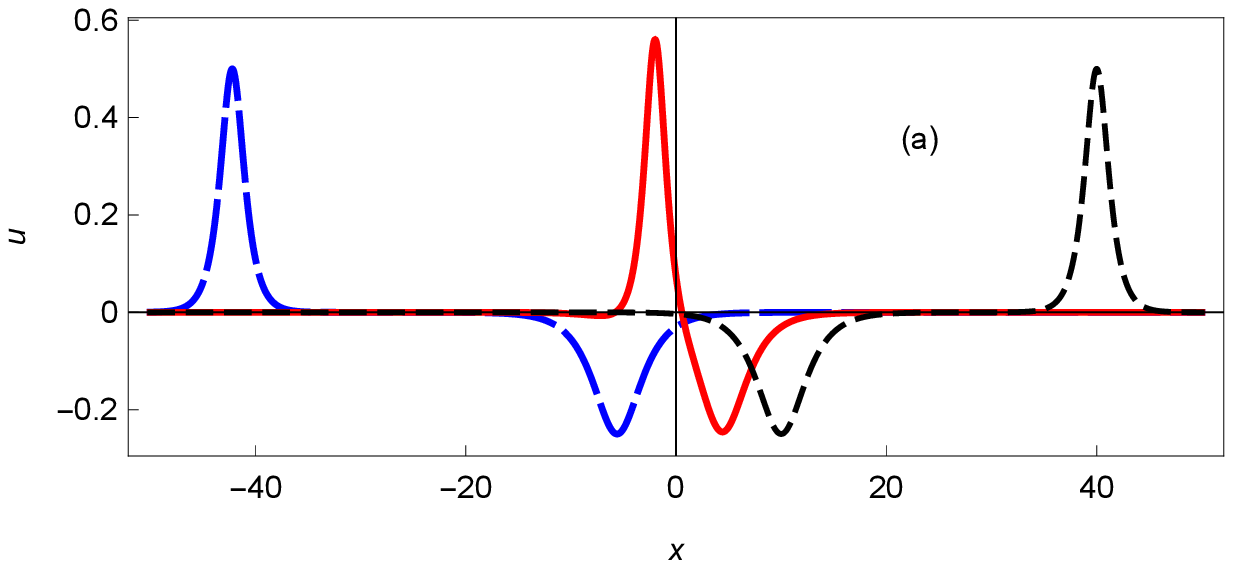}}
\centerline{
\includegraphics[width=100mm]{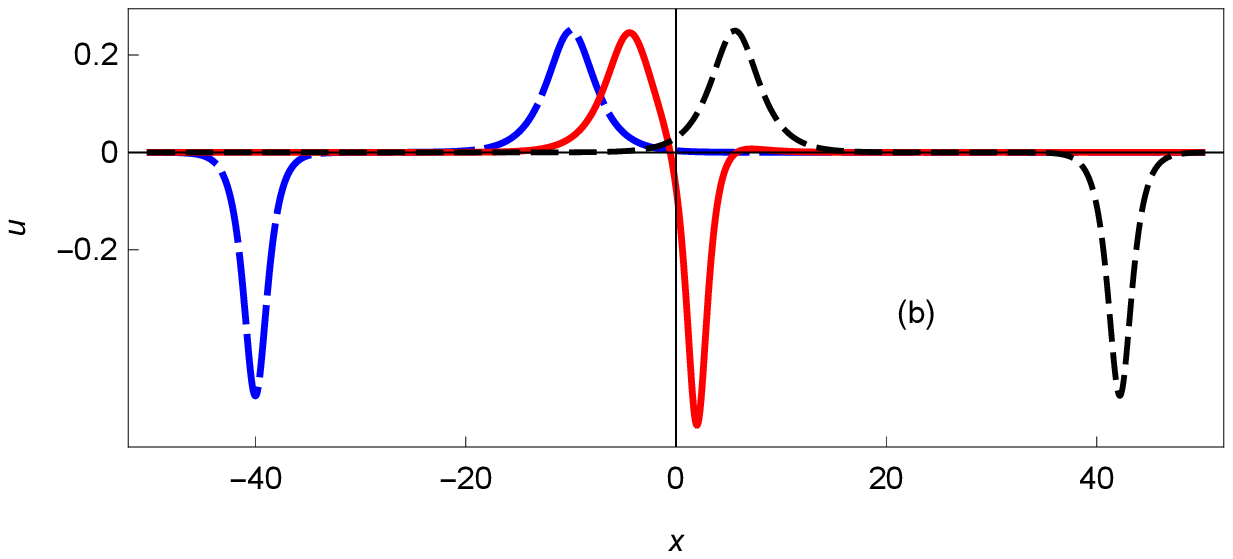}}
\caption{
Part (a) shows the typical overtaking of a smaller hole by a larger hump,
for $k_1=-0.5$ and $k_2=1$, whereas part (b) concerns the opposite, where a
larger hole overtakes a smaller hump, for $k_1=0.5$ and $k_2=-1$.
The curve coding is as in Figure \ref{CC}.}
\label{CR}
\end{figure}

\begin{figure}
\centerline{
\includegraphics[width=60mm]{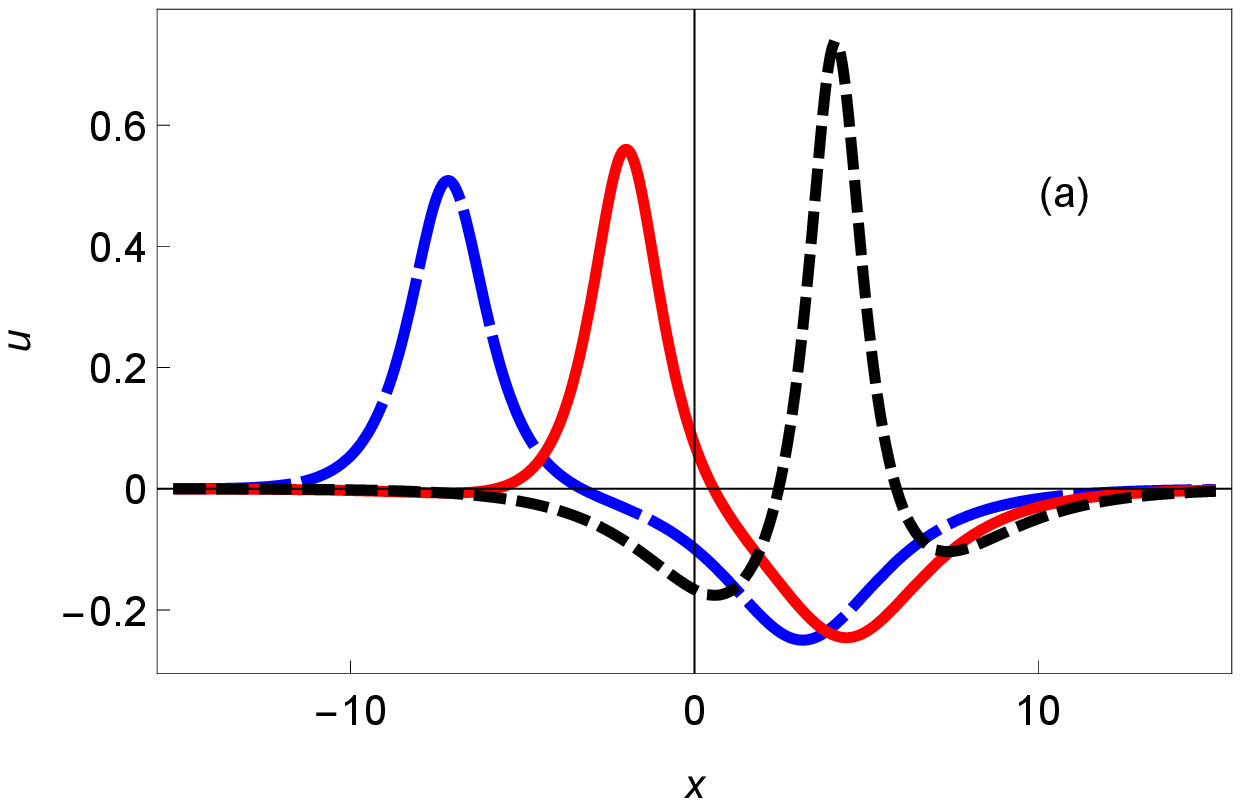} \qquad
\includegraphics[width=60mm]{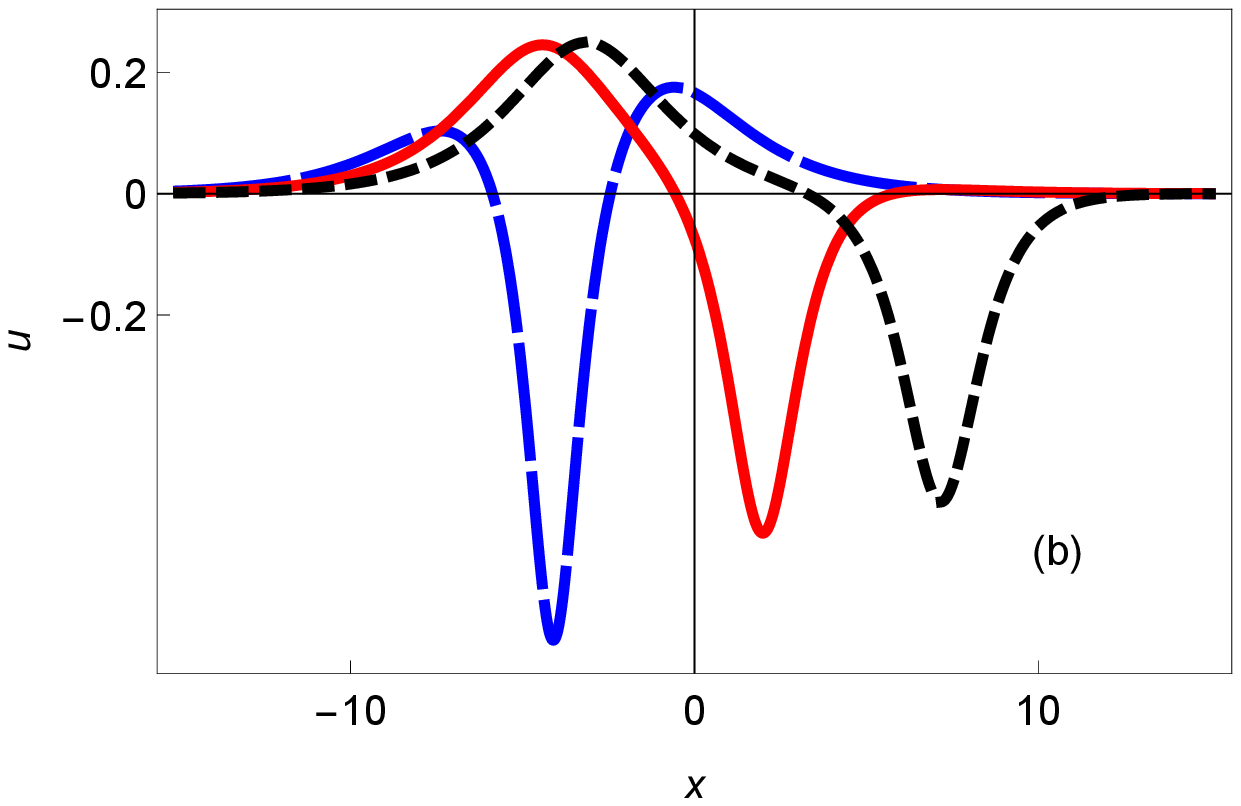}}
\caption{
Details of the centre regions for the overtaking interaction, in part (a) of a
smaller hole by a larger hump, for the parameters given in Figure \ref{CR}(a),
and in (b) of a smaller hump by a larger hole, for the parameters given in
Figure \ref{CR}(b).
The curve coding is as in Figure \ref{CCcent}.}
\label{CRcent}
\end{figure}

The case of two solitons having opposite polarities leads to graphs which have
not been systematically studied before in a plasma physics context.
Figure \ref{CR} illustrates the dynamics of solitons of the opposite polarity
(hump and hole).
Figure \ref{CRcent} gives a clearer (zoomed) view of what happens in the
interaction zones.

We first illustrate in Figure \ref{CR}(a) how a faster, larger but narrower
hump of positive polarity overtakes a slower, smaller but wider hole of
negative polarity.
Larger and smaller refer to the absolute values of the amplitudes of the
respective solitons.
Figure \ref{CR}(b) shows a larger hole overtaking a smaller hump.
The parameters are $k_1=-0.5$, $k_2=1$ in (a), and $k_1=0.5$, $k_2=-1$ in (b).
In part (a), increasing $k_2$ makes the hump larger but narrower; decreasing
$|k_1|$ makes the hole shallower but wider.
For part (b) the converse holds, increasing $k_2$ makes the hole deeper but
narrower; decreasing $|k_1|$ makes the hump smaller but wider.

By reducing the time step near the overtaking region of the two solitons, we
show in Figure \ref{CRcent} how the interaction evolves between the larger and
the smaller soliton.
The parameters are as in Figure \ref{CR}, but with smaller time steps, for
$t=-5, 0, 5$.

In particular, note that in part (a) the hole is split in two when the hump
passes through, and the holes are less deep, rendering the hump larger.
The net result is that the hole is transferred from being ahead of the hump to
trailing it.
On the other hand, in case (b) the hump is split when the hole passes through,
with similar remarks about smaller humps and a deeper hole.

This mechanism has been observed in a variety of other parameter ranges we have
investigated, but omitted here to avoid repetitive figures that are
qualitatively similar.
And, as before, the mechanism occurs for all ratios $|k_1/k_2|$.

The analogy between parts (a) and (b) of Figure \ref{CR} is due to an
underlying antisymmetry:
\beq
u(-x,-t) \big|_{k_1\to - k_1,\,k_2\to - k_2} = - u(x,t).
\eeq
A consequence is that one can prove mathematically from \eqref{umKdVHir} that
$k_1+k_2\to 0$ yields $u(x,t)\to 0$.
This is to be anticipated: When the hole is as deep as the hump is tall, both
will cancel and travel at the same speed, hence one is reduced to the trivial
solution of the mKdV equation.

\section{Electric field profiles}

We recall that $u=\varphi_1$ represents the electrostatic potential.
In the theoretical analysis of acoustic solitons in multispecies plasmas this
is usually the more important quantity.
However, its derivatives have a physical meaning: up to a change of sign, the
first derivative gives the electric field, and the second, the global charge
density of the plasma.
This is in contrast to the original application of the KdV equation --
describing solitary waves on the surface of shallow water -- where the height
of the wave is important but the derivatives of that elevation are not of
physical interest.

\begin{figure}
\centerline{
\includegraphics[width=100mm]{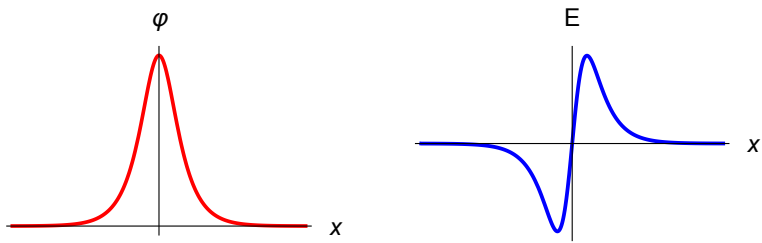}}
\caption{
Generic profiles for mKdV single-soliton solution and associated electric
field.}
\label{SolElec}
\end{figure}

In plasma applications, the electric field of the wave is given by
\begin{equation}
E = - \pd{\varphi}{x}.
\end{equation}
With respect to solitary waves having a potential hump or hole, it follows that
the associated electric fields show a typical bipolar signature, of which a
generic picture is given in Figure \ref{SolElec} for a positive single-soliton
solution \eqref{mKdVSoliton}.
For a potential hump, where the electrostatic potential rises from zero to a
maximum, before decreasing again to zero at infinity, its derivative is
initially positive but then switches to negative.
Because of the minus sign in the definition of the electric field, the bipolar
structure starts with a negative part and ends with a positive part.
For potential holes the opposite occurs.
Thus, the electric field is also an indicator for the polarity of the mode.

The interest in such bipolar electric field structures stems from the fact that
they have been observed in a plethora of space observations by ionospheric and
auroral satellite missions as diverse as GEOTAIL [\cite{Geotail}], POLAR
[\cite{Polar,Franz2005}], FAST [\cite{Fast}] and CLUSTER
[\cite{Pickett2004,Pickett2008,Norgren}] where the electric field (rather than
the electrostatic potential) is observed.
In some of the satellite observations both the electrostatic potential and the
electric field are observed by different instruments [\cite{Fast}], which can
thus serve as a way to check the consistency of the observations.

However, there are many different ways of modelling such structures, which in
the case of electron holes refer to a localized plasma region where the
electron density is lower than the surrounding plasma [\cite{Hutchinson}].
The decreased electron density causes a local maximum in charge density and
electrostatic potential.
They might be viewed as a Bernstein-Greene-Kruskal (BGK) mode [\cite{BGK}], but
can also be described in terms of solitons.
For further details we refer to an excellent overview by \cite{Hutchinson} and
to recent extensions of BGK theory by \cite{Hari2018a,Hari2018b}.
Among the many interpretations, the one that interests us here is that of
electrostatic solitary waves and structures, but there remains ambiguity in
identifying the nature of observed phenomena.

Furthermore, there are many observations that do not fit the simple patterns
and cannot be explained by simple solitons and their electric fields.
A specific example is found in Figure 1(b) in [\cite{Pickett2004}], where
besides the typical bipolar structures there are tripolar structures or
modifications of the bipolar electric field structure by wiggles on the sides.
For the latter we advanced a possible explanation in terms of supersolitons
[\cite{Dubinov2012,Supersolitons,SSAcoustic,Kakad2016,SScollisions}], but in
personal correspondence with the lead author of \cite{Pickett2004} these were
thought to indicate an overtaking or even merging of solitons
[\cite{Pickett2013}].

Moreover, in most of the observed electric fields, the bipolar profiles have
larger or smaller amplitudes, which -- if they are due to propagating solitons
-- implies that the smaller ones will be overtaken by the larger ones.
The property that larger amplitude structures are faster than smaller ones also
holds for large amplitude solitary waves [\cite{DoubleCairns}] which are
commonly described by a Sagdeev pseudopotential formalism [\cite{Sagdeev}].
The drawback of this method is that it usually does not give analytical
expressions for the profiles nor for their interactions.
Outside the KdV range, results will have to come from numerical simulations
(see, e.g., \cite{Kakad2017}).

Of course, observing the precise overtaking of two bipolar structures would be
serendipitous, but signals close to or after an overtaking collision have been
recorded.
Regardless of their theoretical explanation, these observations have been
interpreted in terms of propagating structures, even though that has rarely
been unambiguously confirmed.
There are some CLUSTER observations where a fortuitous configuration of two of
the four satellites was able to capture propagation of the major structures,
while others were already quite distorted [\cite{Pickett2008,Norgren}].

\begin{figure}
\centerline{
\includegraphics[width=100mm]{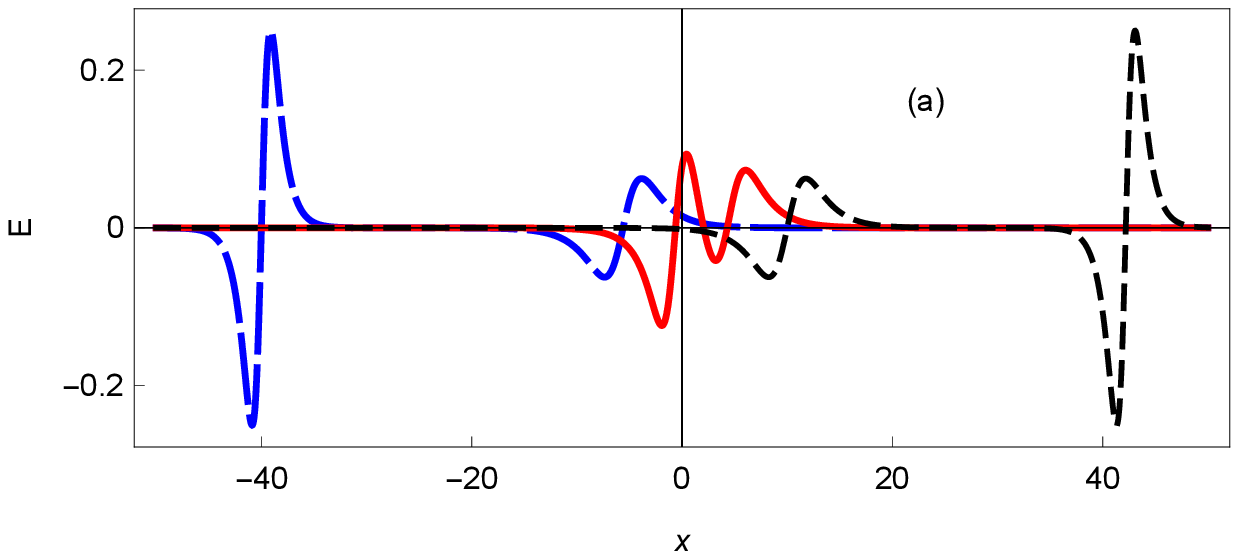}}
\centerline{
\includegraphics[width=100mm]{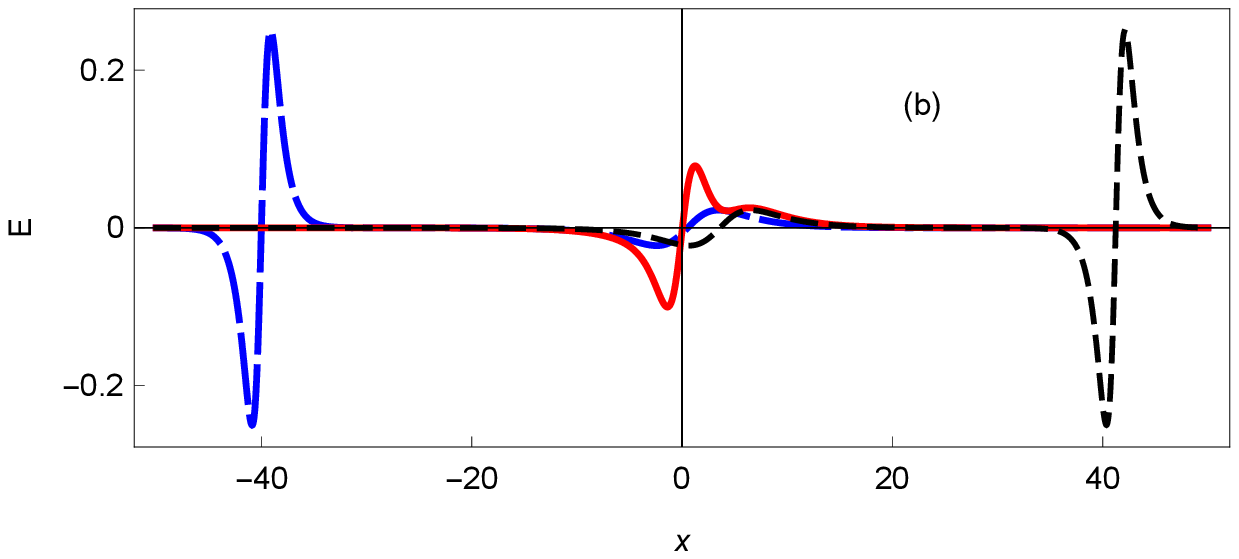}}
\caption{
Electric field profiles of the soliton interactions for (a) $k_1=0.5$ or
(b) $k_1=0.3$ and $k_2=1$, as shown in Figure \ref{CC}.
The curves correspond to $t=-40, 0, 40$, with a similar curve coding.}
\label{ECC}
\end{figure}

Hence, in what follows we focus on the interaction properties of the electric
fields.
The graphs to follow show the negative of the derivative of \eqref{umKdVHir}
with respect to $x$.
We omit the cumbersome mathematical expression of the derivative because it
offers no physical insight.
Using the same parameter conditions as in Figures \ref{CC}--\ref{CRcent},
we start with the positive electrostatic polarities in Figure \ref{ECC}.
As in Figure \ref{CC}, the parameters determining the respective shapes of the
solitons are (a) $k_1=0.5$ and (b) $k_1=0.3$, both with $k_2=1.$
The curves are shown for $t=-40, 0, 40$ with the usual curve coding.
Far away from the interaction region, the electric field shows two
characteristic bipolar signatures, a stronger one for the larger soliton and a
weaker one for the smaller soliton.

\begin{figure}
\centerline{
\includegraphics[width=60mm]{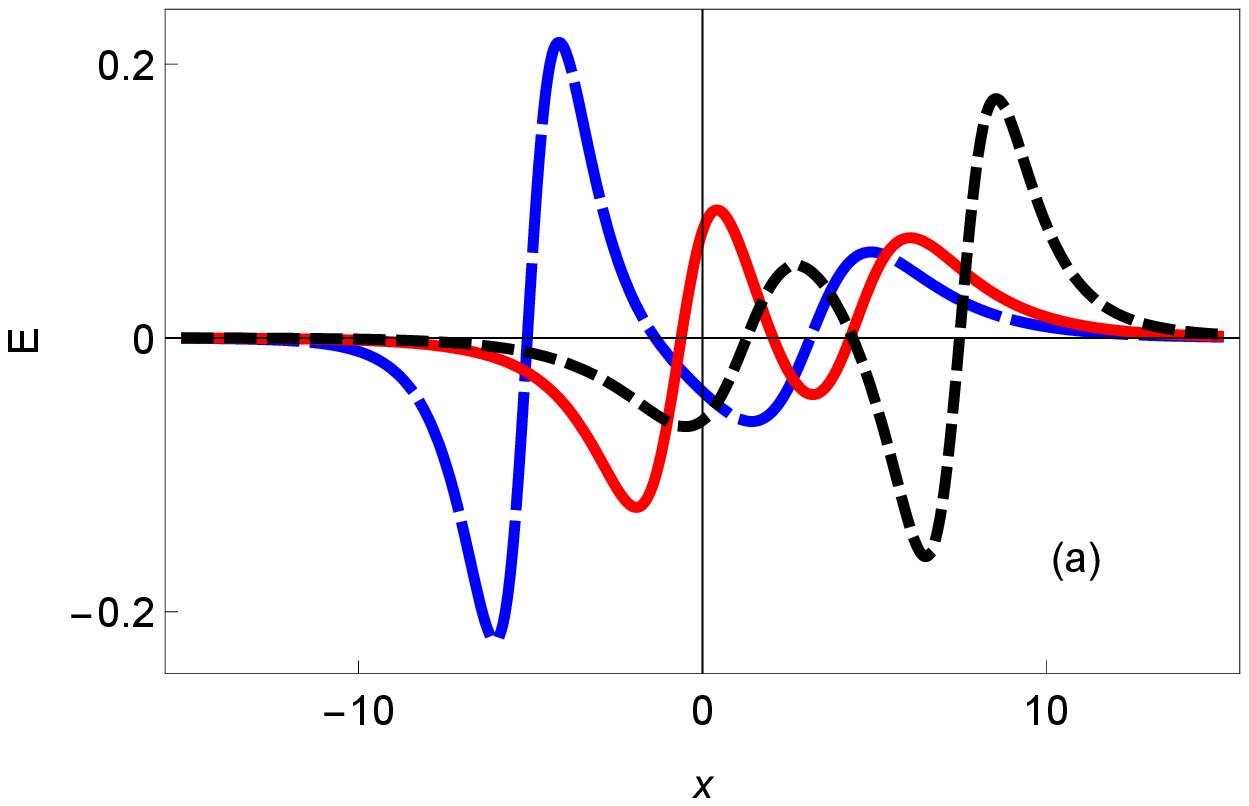} \qquad
\includegraphics[width=60mm]{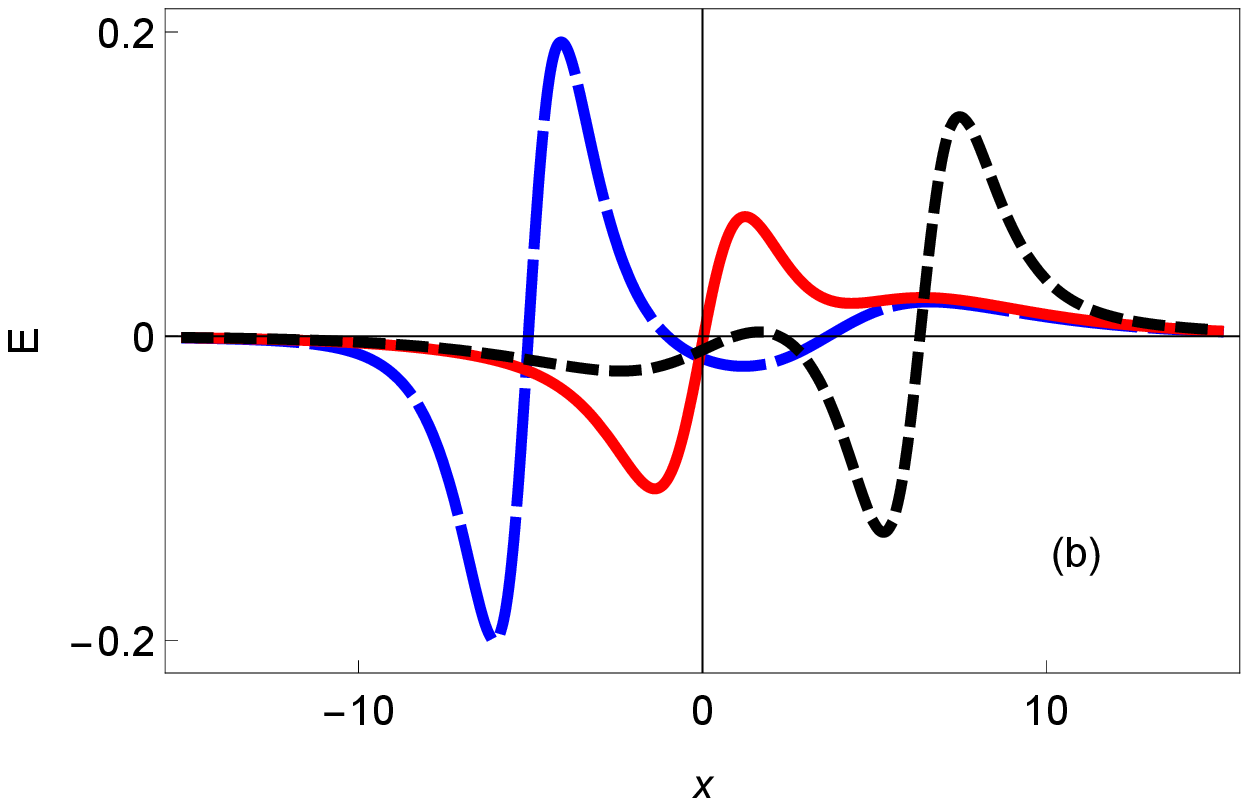}}
\caption{
Details of the electric field profiles for the parameters used in Figure
\ref{ECC}.
The curves correspond to $t=-5, 0, 5$.}
\label{ECCcent}
\end{figure}

As can be seen on these graphs, during the interaction the curves become quite
muddled.
This is shown in more detail in Figure \ref{ECCcent}, for both sets of
parameters.
Parts (a) of Figures \ref{CC} and \ref{CCcent} showed two distinct peaks.
Parts (b) of the same figures illustrated that the peaks briefly merged into a
single distorted one.
Consequently, in Figures \ref{ECC} and \ref{ECCcent} there are always two
distinct bipolar signatures in parts (a), sometimes very close together,
whereas in parts (b) there is a stage when there is only one bipolar structure,
but with wiggles on the wings.
Some of the latter characteristics have also been seen on supersolitons, but in
larger-amplitude descriptions outside the reductive perturbation approach
[\cite{Dubinov2012,Supersolitons}] or in simulations [\cite{Kakad2016}].

\begin{figure}
\centerline{
\includegraphics[width=100mm]{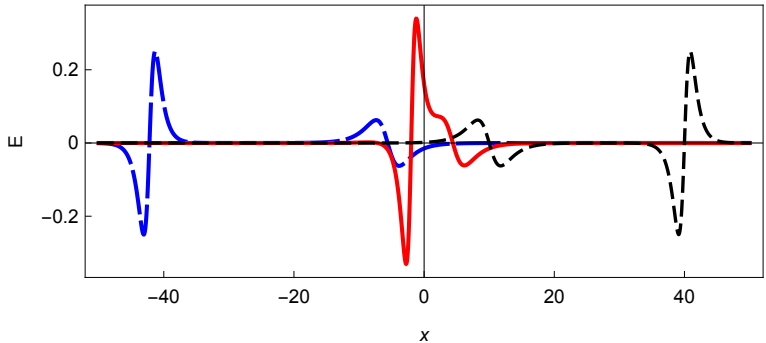}}
\caption{
Electric field profiles for $k_1=-0.5$ and $k_2=1$ for the soliton interactions
shown in Figure \ref{CR}.}
\label{ECR}
\end{figure}

\begin{figure}
\centerline{
\includegraphics[width=70mm]{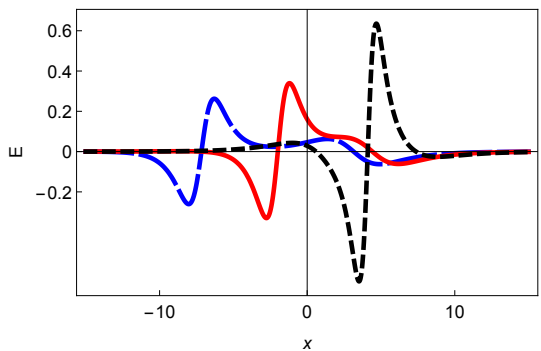}}
\caption{
Details of the electric field profiles for the parameters used in Figure
\ref{ECR}.
The curves correspond to $t=-5, 0, 5$.}
\label{ECRcent}
\end{figure}

Next, in Figures \ref{ECR} and \ref{ECRcent} we graph the overtaking of a
smaller hole by a larger hump.
Here again, the electric field bipolar signatures can be recognized far from
the interaction zone, except that now the signatures differ not only in
amplitude but also in sign.
This is less often seen in space observations but the complicated interaction
region itself might perhaps help with identifying new features.

The figures for the converse case, i.e., the overtaking of a smaller hump by a
larger hole, would be obtained by flipping the graphs shown in Figures
\ref{ECR} and \ref{ECRcent} over the $x$ axis (not shown).

The second derivative of $u=\varphi$ also has a physical interpretation.
Indeed, it follows from Poisson's equation
\beq
\sigma = - \pd{^2\varphi}{x^2}
\eeq
that, up to a sign, it corresponds to the global charge density in the solitary
wave(s).
However, this has rarely been plotted in theoretical papers.
Viewed more as a curiosity than for its additional physical insight, the
available plots are only for single solitons or double layers
[\cite{SSAcoustic}].
Due to the additional local extrema, we have not attempted to graph these
density curves for two interacting mKdV solitons.

\section{Conclusions}

While many papers have covered two-soliton overtaking interactions in plasmas,
they have mostly focused on KdV solitons having the same polarity.
We stress that the KdV equation can only deal with one sign of the quadratic
nonlinearity for a given model.
Likewise, mKdV equations can address the overtaking of same polarity solitons,
in which case profiles and interaction regions look quite similar to what one
obtains for KdV solitons.
Depending on the relative amplitudes (or equivalently, velocities), the salient
feature for two-soliton interaction of the same polarity is that there are
either always two distinct humps or that they briefly merge in the interaction
region into a single distorted hump.
In either case, the collisions are accompanied by phase shifts.

The most interesting case is the overtaking of mKdV solitons of opposite
polarities, where in the interaction region the slower soliton slows down and
splits, to let the faster soliton pass through.
While this happens, the faster soliton is temporarily slightly accelerated.
Far from the interaction region, on either side, the solitons are well
separated and each matches the sech-profile of a single mKdV soliton.

Relevant to plasma physics is the study of the associated electric fields,
which are often seen in ionospheric and magnetospheric satellite observations.
Seemingly, the electric fields are more easily observed than the electrostatic
solitons themselves.
Such electric fields are characterized by typical bipolar structures or
variations thereof.
In this paper we discussed and illustrated the overtaking and interaction
properties of some of these signatures.
As far as we know, this is a novel application.

The case of opposite polarity modes overtaking each other is more rare.
It might therefore be interesting to search for some of these in more
complicated scenarios, in particular, where not all signatures are purely
bipolar and well separated.


\end{document}